\documentclass{bmcart}

	\usepackage{lineno}
	\usepackage{cite}
	\usepackage{etex}
	\usepackage{float}
	\usepackage{url}
	\usepackage{amsmath}
	\usepackage{color}
	\usepackage{listings}
	\usepackage{graphicx}
	\usepackage{amsthm}
	
	\usepackage{amssymb}
	\usepackage{pgfplots}
	\usepgfplotslibrary{groupplots}
	\usepackage{mathtools}
	\usepackage{makeidx}
	\usepackage{amsfonts}
	\usepackage[ansinew]{inputenc}
	\usepackage[usenames,dvipsnames]{pstricks}
	\usepackage{subfigure}
	\usepackage{epsfig}
	\usepackage{pst-grad} 
	\usepackage{pst-plot} 
	\usepackage[colorlinks,hyperindex]{hyperref}
	\makeatletter

\newcommand{\Rmnum}[1]{\expandafter\@slowromancap\romannumeral #1@}
\makeatother

\def\includegraphics{}

\startlocaldefs
\endlocaldefs

\begin{document}

\begin{frontmatter}

\begin{fmbox}
\dochead{Opinion Note}


\title{The Machine that Builds Itself:\\ How the Strengths of Lisp Family Languages Facilitate Building Complex and Flexible Bioinformatic Models}


\author[
   addressref={aff1},                   
   corref={aff1},                       
   email={b.khomtchouk@med.miami.edu}   
]{\inits{BK}\fnm{Bohdan B.} \snm{Khomtchouk}}
\author[
   addressref={aff2},                   
   email={edmund.weitz@haw-hamburg.de}   
]{\inits{EW}\fnm{Edmund} \snm{Weitz}}
\author[
   addressref={aff1},
   email={cwahlestedt@med.miami.edu}
]{\inits{CW}\fnm{Claes} \snm{Wahlestedt}}


\address[id=aff1]{
  \orgname{Center for Therapeutic Innovation and Department of Psychiatry and Behavioral Sciences, University of Miami Miller School of Medicine}, 
   \street{1120 NW 14th ST},  
  \city{Miami, FL},                              
  \cny{USA}      \postcode{33136}
}
\address[id=aff2]{
  \orgname{Hamburg University of Applied Sciences}, 
  \street{Finkenau 35},  
  \city{22081 Hamburg},                              
  \cny{Germany}      
}



\end{fmbox}


\begin{abstractbox}

\begin{abstract}

    We address the need for expanding the presence of the Lisp family
    of programming languages in bioinformatics and computational
    biology research.  Languages of this family, like Common Lisp,
    Scheme, or Clojure, facilitate the creation of powerful and
    flexible software models that are required for complex and rapidly
    evolving domains like biology.  We will point out several
    important key features that distinguish languages of the Lisp
    family from other programming languages and we will explain how
    these features can aid researchers in becoming more productive and
    creating better code.  We will also show how these features make these languages ideal
    tools for artificial intelligence and machine learning
    applications.  We will specifically stress the advantages of
    domain-specific languages (DSL): languages which are specialized
    to a particular area and thus not only facilitate easier research
    problem formulation, but also aid in the establishment of
    standards and best programming practices as applied to the
    specific research field at hand.  DSLs are particularly easy to
    build in Common Lisp, the most comprehensive Lisp dialect, which
    is commonly referred to as the ``programmable programming
    language."  We are convinced that Lisp grants programmers
    unprecedented power to build increasingly sophisticated artificial
    intelligence systems that may ultimately transform machine
    learning and AI research in bioinformatics and computational
    biology.

\end{abstract}




\end{abstractbox}
%

\end{frontmatter}





\section*{Introduction and Background}

The programming language Lisp is credited for pioneering fundamental computer science (CS) concepts that have influenced the development of nearly every modern programming language to date.  Concepts such as tree data structures, automatic storage management, dynamic typing, conditionals, exception handling, higher-order functions, recursion, and more, have all shaped the foundations of today's software engineering community.  The name Lisp derives from ``List processor" \cite{art}, since linked lists are one of Lisp's major data structures, and Lisp source code is comprised of lists.  As such, genome sequence analysis programs can be written naturally through the many convenient Lisp functions available for manipulating list data.  Lists, which are a generalization of graphs, are extraordinarily well supported by Lisp.  As such, programs that analyze sequence data (such as genomics), graph knowledge (such as pathways), and tabular data (such as that handled by R \cite{R}), can be written easily, and can be made to work together naturally in Lisp.  As a programming language, Lisp supports many different programming paradigms each of which can be employed exclusively or intermixed with others; this includes functional and procedural programming, object orientation, meta programming, and reflection.  In the case of the latter, Lisp's reflectivity allows the computer program to examine, introspect, and modify its own structure and behavior at runtime, making it ideal for artificial intelligence and machine learning applications \cite{norvig}.   

In bioinformatics and computational biology, Lisp has successfully been applied to research in systems biology \cite{biobike, biolingua}, database curation \cite{ecocyc1, ecocyc2}, drug discovery \cite{MDL}, network and pathway -omics analysis \cite{sri1, sri2, sri3, sri4, sri5}, single nucleotide polymorphism analysis \cite{snper1, snper2, snper3}, and RNA structure prediction \cite{structurelab1, structurelab2, structurelab3}.  In general, the Lisp family of programming languages (LFLs), which includes Common Lisp, Scheme, and Clojure, has powered multiple applications across fields as diverse as \cite{franz}: animation and graphics, AI, bioinformatics, B2B and e-commerce, data mining, electronic design automation/semiconductor applications, embedded systems, expert systems, finance, intelligent agents, knowledge management, mechanical computer-aided design (CAD), modeling and simulation, natural language, optimization, risk analysis, scheduling, telecommunications, and web authoring. 

Programmers often test a language's mettle by how successfully it has fared in commercial settings, where big money is often on the line.  To this end, Lisp has been successfully adopted by commerical vendors such as the Roomba vacuuming robot \cite{PCL, L}, Viaweb (acquired by Yahoo! Store) \cite{viaweb}, ITA Software (acquired by Google Inc. and in use at Orbitz, Bing Travel, United Airlines, US Airways, etc) \cite{orbitz}, Mirai (used to model the Gollum character for the Lord of the Rings movies) \cite{mirai}, Boeing \cite{boeing}, AutoCAD \cite{autocad1}, among others.  Lisp has also been the driving force behind open source applications like Emacs \cite{emacs} and Maxima \cite{maxima}, which both have existed for decades and continue to be used worldwide.   

Amongst the LFLs, Common Lisp has been described as the most powerful and accessible modern language for advanced biomedical concept representation and manipulation \cite{ira}.  Scheme \cite{scheme} is an elegant and compact subset of Common Lisp that supports a minimalistic core language and an excellent suite of language extensions tools.  However, Scheme has traditionally mainly been used in teaching and CS research and its implementors have thus prioritized small size, the functional programming paradigm, and a certain kind of ``cleanliness" over more pragmatic features.  As such, Scheme is considered far less popular than Common Lisp for building large-scale applications \cite{PCL}.  

The third most common LFL, Clojure \cite{clojure, hickey}, is a rising star language in the modern software development community.  Clojure specializes in the parallel processing of big data through the Java Virtual Machine (JVM), recently making its debut in bioinformatics and computational biology research \cite{bioclojure, cljam, GATKclojure}.  Most recently, Clojure was used to parallelize the processing and analysis of SAM/BAM files \cite{cljam}.  Furthermore, the BioClojure project provides seeds for the bioinformatics community that can be used as building blocks for writing LFL applications.  As of now, BioClojure consists of parsers for various kinds of file formats (UniProtXML, Genbank XML, FASTA, and FASTQ), as well as wrappers of select data analysis programs (BLAST, SignalP, TMHMM, and InterProScan) \cite{cljam}.  Such projects may also soon find their way into bioinformatics and computational biology applications.

As a whole, Lisp continues to develop new offshoots.  A relatively recent addition to the family is Julia \cite{julia}.  Although it is sometimes touted ``C for scientists" and caters to a different community due to its syntactical proximity to Python, it is a Lisp at heart and certainly worth watching.

\section*{Macros and domain-specific languages}

Lisp is a so-called \textit{homoiconic} language, which means that Lisp code is
represented as a data structure of the language itself.  In more technical terms, a Lisp program has direct access to (and can modify) its abstract syntax tree.  This property enables Lisp to have a macro system that remains undisputed in the programming language world \cite{SO}.  While ``macros" in languages like C have the same name, they are essentially just text substitutions performed on the source code before it is compiled and they can't always reliably preserve the lexical structure of the code.  Lisp macros, on the other hand, operate at the syntactic level.  They transform the program structure itself and, as opposed to C macros, are written in the same language they work on and have the full language available all the time.  Lisp macros are thus not only used for moderately simple ``find and replace" chores, but can apply extensive structural changes to a program.  This includes tasks that are impossible in other languages, like the introduction of new control structures or pattern matching capabilities or the integration of code with markup languages \cite{clmarkup, clwho}.  In addition to that, Common Lisp even offers access to its ``reader" which means that code can be manipulated (in Lisp) before it is parsed \cite{weitz}.  This enables Lisp programs to completely change their surface syntax if necessary \cite{xmlisp, clinterpol, infix}.

These features make Lisp an ideal tool for the creation of \textit{domain-specific languages}: languages that are custom-tailored to a specific problem domain but can still have access to all of Lisp.  A striking example is Common Prolog \cite{commonprolog}, a professional Prolog system implemented and embedded in Common Lisp.  In bioinformatics, the Biolingua \cite{biolingua} project (now called BioBIKE) built a cloud-based general symbolic biocomputing DSL entirely in Common Lisp. The system, which could be programmed entirely through the browser, was its own complete biocomputing language, which included a built-in deductive reasoner, called BioDeducta \cite{shrager}. Biolingua programs, guided by the reasoner, would invisibly call tools such as Blast \cite{blast} and Bioconductor \cite{bioconductor} on the server-side, as needed.  Symbolic biocomputing has also previously been used to create user-friendly visual tools for interactive data analysis and exploration \cite{travers}.

\section*{Other Unique Strengths}

In addition to homoiconicity, Lisp has several other features which set it apart from mainstream languages:

\begin{itemize}
\item In Lisp, programmers usually work in a special incremental interactive programming environment called the read-eval-print loop (REPL) \cite{repl, repl2}.  The REPL enables a paradigm that allows the programmer to continually interact with their program as it is developed.  This is similar to the way Smalltalk ``images" evolve \cite{weitz} and very different from the usual edit-compile-link-execute cycle of C-like languages.  This approach lends itself very well to explorative programming and rapid prototyping.  The REPL enables the programmer to write a function, test it, change it, try a different approach, etc., while never having to stop for any lengthy compilation cycles \cite{PCL}.
\item Common Lisp was designed from the ground up to create large, complex, and long-running applications and thus supports software ``hot swapping": the code of a running program can be changed without the need to interrupt it.  This includes features like the ability of CLOS (Common Lisp's object system) to change the classes of existing objects.
\item Lisp invented exception handling, and Common Lisp in particular has an error handling facility (the ``condition system" \cite{PCL}) that goes far beyond most other languages: it doesn't necessarily unwind the stack if an exception occurs and instead offers so-called restarts to programmatically continue ``where the error happened."  This system makes it easy to write robust software, which is an essential ingredient to building industry-strength fault-tolerant systems capable of handling a variety of conditions, a trait especially useful for artificial intelligence and machine learning applications.
\item Common Lisp implementations usually come with a sophisticated ``foreign function interface" (FFI) \cite{PCL} which allows direct access from Lisp to code written in C or C++ and sometimes also to Java code.  This enables Lisp programmers to make use of libraries written in other languages, making those libraries a direct strength of Lisp.  For instance, it is simple to call Bioconductor from Lisp, just as Python and other programming languages do \cite{gautier, prins}.  Likewise, Clojure runs on the Java Virtual Machine and, thus, has immediate access to all of Java's libraries.
\end{itemize}

It has been shown that these features, together with other amenities like powerful debugging tools that Lisp programmers take for granted, offer a significant productivity boost to programmers \cite{gat, norviglisp}.  Lisp also gives programmers the ability to implement complex data operations and mathematical constructs in an expressive and natural idiom \cite{fenwick}.  

As a whole, LFLs are known to attract a strong audience \cite{myth, bipolar} and to be popular with many historical CS figures and other prolific and extremely productive programmers.  Much speculation has arisen to explain this phenomenon, namely why ``super programmers'' are so drawn to and cater to Lisp, but so far the results have been inconclusive and generally interspersed across website posts, blogs, and miscellaneous comment sections.

\section*{Speed considerations}

The interactivity and flexibility of Lisp languages is something that can usually only be found (if at all) in interpreted languages.  This might be the origin of the old myth that Lisp is interpreted and must thus be slow.  But this is not true.  Compilers for Lisp have existed since 1959 and all major Common Lisp implementations nowadays can compile directly to machine code which is often on par with C code \cite{verna} or only slightly slower.  (Some also offer an interpreter in addition to the compiler, but examples like Clozure Common Lisp demonstrate that you can have a compiler-only Common Lisp.)  For example, CL-PPCRE, a regular expression library written in Common Lisp, runs faster than Perl's regular expression engine on some benchmarks, even though Perl's engine is written in highly tuned C \cite{PCL}.

While programmers who use interpreted languages like Python or Perl for their convenience and flexibility will have to resort to writing in C for time-critical portions of their code, Lisp programmers can usually have their cake and eat it too.  And not only will the code created by Lisp compilers be quite efficient by default, Common Lisp in particular offers unique features to optimize those parts of the code (usually only a tiny fraction) which really need to be as fast as possible \cite{weitz}.  This includes so-called compiler macros, which can transform function calls into more efficient code at runtime, and a mandatory disassembler which enables programmers to fine-tune time-critical functions until the compiled code matches their expectations.

It should also be emphasized that while the C or Java compiler is "history" once the compiled program is started, the Lisp compiler is always present and can thus generate new, fast code while the program is already running.

To further debunk the popular misconception that Lisp languages are slow, Clojure was used to process and analyze SAM/BAM files \cite{cljam} with significantly less lines of code and almost identical speeds as SAMTools \cite{samtools}, which is written in the C programming language.

\section*{Rewards and Challenges}

In general, early adopters of a language framework are better poised to reap the scientific benefits, as they are the first to set out building the critical libraries, ultimately attracting and retaining a growing share of the research and developer community.  Since library support for bioinformatics tasks in the Lisp family of programming languages (Clojure, Common Lisp, and Scheme) is yet in its early stages and on the rise, and there is (as of yet) no officially established bioinformatics Lisp community, there is plenty of opportunity for high-impact work in this direction.

It is well-known that the best language to choose from should be the one that is best suited to the job at hand.  Yet, in practice, few programmers may consider a non-mainstream programming language for a project, unless it offers strong, community-tested benefits over its popular contenders for the specific task at hand.  Often times, the choice comes down to library support: does language X already offer well-written, optimized code to help solve my research problem, as opposed to language Y (or perhaps language Z)?  In general, new language adoption boils down to a chicken-and-egg problem: without a large user-base it is difficult to create and maintain large-scale, reproducible tools and libraries.  But without these tools and libraries, there can never be a large user-base. Hence, a new language must have a big advantage over the existing ones and/or a powerful corporate sponsorship behind it to compete \cite{java}.  Most often, a positive feedback loop is generated by repositories of useful libraries attracting users, who, in turn, add more functional libraries, thereby raising a programming language's popularity, rather than reflecting its theoretical potential.    

With mainstream languages like R \cite{R} and Python \cite{Python} dominating the bioinformatics and computational biology scene for years, large-scale software development and community support for other less popular language frameworks has weened to relative obscurity.  Consequently, languages winning over increasingly growing proportions of a steadily expanding user-base have the effect of shaping research paradigms and influencing modern research trends.  For example, R programming generally promotes research that frequently leads to the deployment of R packages to Bioconductor \cite{bioconductor}, which has steadily grown into the largest bioinformatics package ecosystem in the world, whose package count is considerably ahead of BioPython \cite{biopython}, BioClojure \cite{bioclojure}, BioPerl \cite{bioperl}, BioJava \cite{biojava}, BioRuby \cite{bioruby}, BioJulia \cite{biojulia}, or SCABIO \cite{scabio}.  As a community repository of bioinformatics packages, BioLisp does not yet exist as such, albeit its name currently denotes the native language of BioBIKE \cite{biobike, shrager}, a large-scale bioinformatics Lisp application.  

Likewise, given the choice, R programmers interested in deploying large-scale applications are more likely to branch out to releasing web applications (e.g., Shiny \cite{shiny}) than to GUI binary executables, which are generally more popular with lower-level languages like C/C++ \cite{C++}.  As such, language often dictates research direction, output, and funding.  Questions like ``who will be able to read my code?", ``is it portable?", ``does it already have a library for that?", or ``can I hire someone?" are pressing questions, often inexorably shaping the course and productivity of a project.           

\section*{Perspectives and Outlook}

Historically speaking, Lisp is the second oldest (second only to Fortran) programming language still in use and has influenced nearly every major programming language to date with its constructs \cite{graham}.  For example, it may be surprising to learn that R is written atop of Scheme \cite{ihaka}.  In fact, R borrows directly from its Lisp roots for creating embedded domain specific languages within R's core language set \cite{advancedR}.  For instance, ggplot2 \cite{ggplot2}, dplyr \cite{dplyr}, and plyr \cite{plyr} are all examples of DSLs in R.  This highlights the importance and relevance of Lisp as a programmable programming language, namely the ability to be user-extensible beyond the core language set.  Given the wide spectrum of domains and subdomains in bioinformatics and computational biology research, it follows that similar applications tailored to genomics, proteomics, metabolomics, or other research fields may also be developed as extensible macros in Common Lisp.  By way of analogy, perhaps a genomics equivalent of ggplot2 or dplyr is in store in the not-so-distant future.  Advice for when such pursuits are useful is readily available \cite{mernik}.


\section*{Conclusions}

New programming language adoption in a scientific community is both a challenging and rewarding process.  Here we advocate for and propose a greater inclusion of the Lisp family of programming languages (LFLs) into large-scale bioinformatics research, outlining the benefits and opportunities of the adoption process.  We provide historical perspective on the influence of language choice on research trends and community standards, and emphasize Lisp's unparalleled support for homoiconicity, domain-specific languages, extensible macros, and error handling, as well as their significance to future bioinformatics research.  We forecast that the current state of Lisp research in bioinformatics and computational biology is highly conducive to a timely establishment of robust community standards and support centered around not only the development of bioinformatic domain-specific libraries, but also the rise of highly customizable and efficient machine learning and AI applications written in languages like Common Lisp, Clojure, and Scheme. 

\section*{Key Points}
\begin{itemize}
\item Lisp's treatment of code as data and data as code (a property called homoiconicity), as well as the ability of Lisp programs to examine, introspect, and modify their own structure and behavior at runtime (a property called reflectivity), permits flexible software engineering practices that are conducive to producing unparalleled AI and machine learning applications in bioinformatics and computational biology.
\item The Lisp family of programming languages (Common Lisp, Scheme, Clojure) makes it easy to create extensible macros, which facilitate the creation of modularized extensions to help bioinformaticians easily create plug-ins for their software.  This, in turn, paves the way for creating enterprise-level, fault-tolerant domain-specific languages in any research area or specialization.
\item The current state of Lisp research in bioinformatics and computational biology is at a point where an official BioLisp community is likely to be established soon.  
\end{itemize}


\begin{backmatter}

\section*{Author information}
  Center for Therapeutic Innovation and Department of Psychiatry and Behavioral Sciences, University of Miami Miller School of Medicine, 1120 NW 14th ST, Miami, FL, USA 33136.  Correspondence: \url{b.khomtchouk@med.miami.edu}.

\section*{Competing interests}
  The authors declare that they have no competing interests.

\section*{Author's contributions}
  BBK conceived the study.  BBK and EW wrote the paper.  BBK, EW, and CW planned the study.  All authors read and approved the final manuscript.  

\section*{Acknowledgements}
BBK dedicates this work to the memory of his uncle, Taras Khomchuk.  BBK wishes to acknowledge the financial support of the United States Department of Defense (DoD) through the National Defense Science and Engineering Graduate Fellowship (NDSEG) Program: this research was conducted with Government support under and awarded by DoD, Army Research Office (ARO), National Defense Science and Engineering Graduate (NDSEG) Fellowship, 32 CFR 168a.  CW thanks Jeff Shrager for critical review and helpful comments on the manuscript.


\bibliography{lisp_bioinformatics}
\bibliographystyle{unsrt}








\section*{Abbreviations used}
  LFL: Lisp family of languages \\
  DSL: domain specific language(s) \\
  AI: artificial intelligence \\
  CS: computer science \\
  PCA: principal component analysis \\
  CAD: computer-aided design \\
  JVM: Java Virtual Machine \\
  API: application programming interface \\
  REPL: read-eval-print loop \\
  CLOS: Common Lisp Object System \\
  FFI: Foreign Function Interface \\
  REPL: read-eval-print loop

\end{backmatter}

\end{document}